\documentclass[epj]{svjour}
\usepackage{graphics}
\usepackage{graphicx}
\begin{document}
\title{
Formation of the $S=-1$ resonance $X(2265)$\\
 in the reaction $pp \rightarrow X + K^+$ at 2.50 and 2.85 GeV
}
\subtitle{June 7, 2012}
\author{
 P.~Kienle\inst{1,2} 
 \and
 M.~Maggiora\inst{3}
 \and
 K.~Suzuki\inst{2}
 \and
 T.~Yamazaki\inst{4,5}
 \and
 M.~Alexeev\inst{3,14}
 \and
 F.~Balestra\inst{3},
 \and
 Y.~Bedfer\inst{6} 
 \and
 R.~Bertini\inst{3,6}
 \and
L.~C.~Bland\inst{7}, 
 \and
A.~Brenschede\inst{8} 
 \and
F.~Brochard\inst{6}
 \and
M.~P.~Bussa\inst{3} 
 \and
M. Chiosso\inst{3}
 \and
Seonho~Choi\inst{7} 
 \and
M.~L.~Colantoni\inst{3}
 \and
R.~Dressler\inst{13} 
 \and
M.~Dzemidzic\inst{7} 
 \and
J.-Cl.~Faivre\inst{6}
 \and
A. Ferrero\inst{3}
 \and
L.~Ferrero\inst{3} 
 \and
J.~Foryciarz\inst{10,11} 
 \and
I. Fr\"ohlich\inst{8} 
 \and
V.~Frolov\inst{9} 
 \and
R.~Garfagnini\inst{3} 
 \and
A.~Grasso\inst{3}
 \and
S.~Heinz\inst{3,6} 
 \and
W.~W.~Jacobs\inst{7} 
 \and
W.~K\"uhn\inst{8} 
 \and
A.~Maggiora\inst{3}
 \and
D.~Panzieri\inst{12} 
 \and
H.-W.~Pfaff\inst{8}
 \and
G.~Pontecorvo\inst{3,9} 
 \and
A.~Popov\inst{9}
 \and
J.~Ritman\inst{8} 
 \and
P.~Salabura\inst{10} 
 \and
 \and
V.~Tchalyshev\inst{9}
 \and
F. Tosello\inst{3}
 \and
S.E. Vigdor\inst{7}
 \and
G. Zosi\inst{3} 
}                     
%
%
\institute{
Excellence Cluster Universe, Technische Universit\"at M\"unchen, Garching, Germany
\and
Stefan Meyer Institute for Subatomic Physics, Austrian Academy of Sciences, Vienna, Austria
\and
Dipartimento di Fisica Generale ``A. Avogadro'' and INFN, Torino, Italy
\and
Department of Physics, University of Tokyo, Tokyo, 116-0033 Japan
\and
RIKEN Nishina Center, Wako, Saitama, 351-0198 Japan
\and
Laboratoire National Saturne, CEA Saclay, France
\and
Indiana University Cyclotron Facility, Bloomington, Indiana, U.S.A.
\and
II. Physikalisches Institut, Universit\"at Gie\ss{}en, Germany
\and
JINR, Dubna, Russia
\and
M.~Smoluchowski Institute of Physics, Jagellonian University, Krak\'{o}w, Poland
\and
H. Niewodniczanski Institute of Nuclear Physics, Krak\'ow, Poland
\and
Universit\`{a} del Piemonte Orientale and INFN, Torino, Italy
\and
Forschungszentrum Rossendorf, Germany
\and
INFN, Trieste, Italy
}
\date{Received: date / Revised version: date}
%
\abstract{
Analyzing DISTO data of $pp \rightarrow p \Lambda K^+$ at $T_p = 2.50$ and 2.85 GeV to populate a previously reported $X(2265)$ resonance with $M_X = 2267$ MeV/$c^2$ and $\Gamma_X = 118$ MeV at 2.85 GeV, we found that the production of $X(2265)$ at 2.50 GeV is much less than that at 2.85 GeV (less than 10\%), though it is expected from a kinematical consideration to be produced as much as 33\% of that at 2.85 GeV. The small population of $X(2265)$ at 2.50 GeV is consistent with the very weak production of $\Lambda (1405)$ at the same incident energy toward its production threshold, thus indicating that $\Lambda (1405)$ plays an important role as a doorway state for the formation of $X(2265)$.
} 
\maketitle
%

Recently, analyzing a set of the DISTO data of an exclusive reaction, $pp \rightarrow p \Lambda K^+$, taken at an incident kinetic energy of $T_p$ = 2.85 GeV, we found \cite{Yamazaki:10} a 
broad resonance with a mass of $M_X = 2267 \pm 2(stat) \pm 5(syst)$ MeV/$c^2$ and a width of  $\Gamma_X = 118 \pm 8(stat) \pm 10(syst)$ MeV, 
in the invariant-mass spectrum $M(p \Lambda)$, and 
also in the missing-mass spectrum $\Delta M(K^+)$. For the time being, we call this resonance 
$X(2265)$. 
An indication for a similar resonance in $K^-$ absorption by light nuclei was reported from FINUDA \cite{FINUDA}. 


  For further understanding the nature of $X(2265)$ we studied the entrance-channel behavior of the $pp$ reaction, and analyzed more reaction data from DISTO taken at 2.50 GeV, at which energy the formation of $X(2265)$ should still be kinematically allowed (the nominal threshold energy: $T_p^{\rm thres} (X(2267)) = 2.19$ GeV), whereas the formation of the $\Lambda(1405)$ resonance (abbreviated here as $\Lambda^*$) is expected to become very weak toward its production threshold ($T_p^{\rm thres} (\Lambda^*) = 2.42$ GeV). This will clarify the nature of $X(2265)$ and let us know if the $\Lambda^*$ plays an essential role in the formation process.

\begin{figure}[tb]
  \begin{center}
  \includegraphics{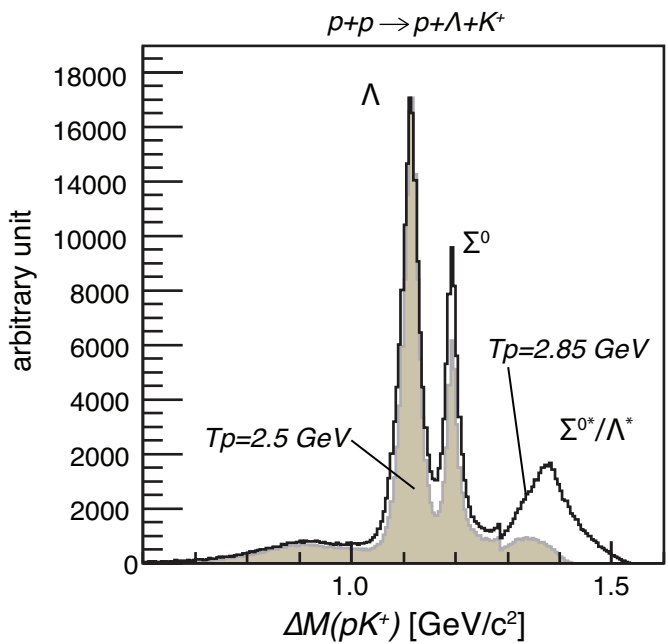}
 \caption{
 Comparison of missing-mass $\Delta M(pK^+)$ spectra  of the $pp \rightarrow p \Lambda K^+$ reaction at $T_p = 2.85$ GeV (solid histogram) and 2.50 GeV (shaded) normalized with the numbers of produced $\Lambda's$. The momentum distributions of the two particles, $p$ and $\Lambda$, are examined to prove that the momentum acceptance for $\Delta M(pK^+)$ is flat.}
\label{fig:DMpK}
 \end{center} 
 \end{figure}

The DISTO experiment was carried out with the SATURNE accelerator at Saclay \cite{Balestra:99,Maggiora:01}. Here, we have analyzed the data set of the exclusive reaction products $p \Lambda K^+$ at $T_p = 2.50$ GeV, and compared the results with those at 2.85 GeV \cite{Yamazaki:10,Maggiora:09} using the same analysis method and checking the acceptance corrections at both incident energies. About 125k exclusive events were selected from a neutral hyperon missing-mass spectrum $\Delta M(pK^+)$, using the previous procedures described in detail in \cite{Yamazaki:10,Maggiora:09}. 

 Figure~\ref{fig:DMpK} shows  
 missing-mass spectra $\Delta M(p K^+)$ at $T_p = 2.85$ and 2.50 GeV, obtained requiring the invariant mass of the $p-\pi^-$ system to be that of the $\Lambda$. The momentum acceptances of the $p$ and $K^+$ are proven to be flat at both incident energies. Thus, the cut-off of the missing-mass spectrum $\Delta M(pK^+)$ at 1.4 GeV/$c^2$ for $T_p = 2.50$ GeV  is shown to be not due to a change of the momentum acceptance of the $p$ and $K^+$. The $\Delta M(p K^+)$ spectrum at 2.85 GeV shows lines at masses of $\Lambda$, $\Sigma ^0$, and $\Sigma ^0(1385)$ ($\equiv {\Sigma^0}^*$) + $\Lambda (1405)$ ($\equiv \Lambda^*$), the latter two being unresolved. 

 Zychor {\it et al.} \cite{Zychor} made an analysis on their ${\Sigma^0}^*+\Lambda^*$ composite peak in $pp \rightarrow p \Lambda K^+$ events at 2.83 GeV, and found that it is composed of  ${\Sigma^0}^*$ and $\Lambda^*$ by an intensity ratio of $I(\Lambda^*) : I({\Sigma^0}^*) = 1.00 : 2.37$. They used the missing-mass information for $\pi^0$ and $\Sigma^0 \rightarrow \Lambda \gamma$ to distinguish between ${\Sigma^0}^* \rightarrow \Lambda \pi^0$ and $\Lambda^* \rightarrow \Sigma^0 \pi^0 \rightarrow \Lambda  \gamma  \pi^0$, and obtained the individual cross sections as $\sigma({\Sigma^0}^*) = 4.0~\mu$b and $\sigma(\Lambda^*) = 4.5~\mu$b. 
  
 The $\Delta M(p K^+)$ spectrum at 2.50 GeV, which is overlaid in Fig.~\ref{fig:DMpK}, shows again the lines for the production of $\Lambda$ and $\Sigma^0$, but  the ${\Sigma^0}^*+ \Lambda^*$ complex peak appears to be very much reduced and shifted toward the lower mass; obviously, the formation of the $\Lambda^*$ resonance is kinematically hindered toward the threshold ($T_p (\Lambda^*) \sim 2.42$ GeV) at an incident energy of 2.50 GeV. The dramatic change of the $\Lambda^*$-resonance shape and intensity at $T_p = 2.50$ GeV is understood by a realistic calculation taking into account  the finite width \cite{Hassanvand:12}. We set an upper limit for the ratio of the cross sections of $\Lambda^*$ at $T_p = 2.50$ GeV to 2.85 GeV to be 0.10. 
 
 
For the analyses of the reaction spectra, we take an uncorrected raw experimental spectral distribution ($RAW^{(\alpha)}$) for a certain kind, $\alpha$, which is a raw distribution not corrected for acceptance, and a corresponding simulated distribution ($SIM^{(\alpha)}$) calculated for events of the three-body reaction $p \Lambda K^+$ assuming a uniform phase-space distribution, folded with the DISTO geometrical acceptance. To avoid possible uncertainties in the acceptance correction, we adopt a {\it deviation spectrum} method to obtain an acceptance-compensated presentation of the spectrum of $\alpha$, by calculating
\begin{equation}
DEV^{(\alpha)} = RAW^{(\alpha)}/SIM^{(\alpha)}
\end{equation}
for all bins. 
 A thus obtained $DEV$ spectrum is not only acceptance compensated, but also is free from dropping phase-space densities (bell-shaped) near their boundaries. A $DEV$ spectrum is in general flat and linear, but will reveal a non-linear structure when a physically meaningful deviation from a uniform phase-space distribution occurs, such as a resonance.


 From the previous analysis we learned that all $p \Lambda K^+$ events are clearly distinguished by their proton angular distribution, which consists of a sharp forward/backward component and a broad large-angle component \cite{Yamazaki:10}. The observed angular distribution of protons is explained by considering the ordinary reaction process, 
 \begin{equation}
p + p \rightarrow p +  \Lambda + K^+,\label{eq:pLK}
\end{equation}
 without invoking resonances. A simple estimation of the angular distribution and the $M(p\Lambda)$ spectrum formulated in Ref. \cite{Akaishi:11} is used here, and explained in what follows. 
  The incident proton with a c.m. momentum of $\vec{p_0}$ produces a scattered proton with momentum $\vec{p_1}$, a $\Lambda$ particle with $\vec{p_2}$ and a $K^+$ with $\vec{p_3}$. The momentum transfer from the incident proton to the scattered proton, $Q = |\vec{p_0}-\vec{p_1}|$ is given by
\begin{equation}
Q^2 = p_1^2 + [\frac{1}{2} + \frac{M_p}{M_{\Lambda} + m_K}]^2 {p_0}^2 
                - 2\, [\frac{1}{2} + \frac{M_p}{M_{\Lambda} + m_K}] p_0 p_1 X_1,
\end{equation}
with $X_1 =  (\hat{p_0} \cdot \hat{p_1})$. The cross section of the process $pp \rightarrow p \Lambda K^+$ is given by a $T$ matrix, which depends on $Q^2$, as
\begin{equation}
 T(Q^2) = V_0 [\frac{1}{1 + b_1^2 Q ^2} + G \frac{1}{1 + b_2^2 Q^2}],
\label{eq:T}\end{equation}
where 
\begin{equation}
b_1 = \frac{\hbar c}{m_B^{(1)}}, ~ b_2 = \frac{\hbar c}{m_B^{(2)}}
\end{equation}
with $m_B^{(1)}$ and $m_B^{(2)}$ being representative intermediate boson masses for small and large momentum transfers, respectively. The observed very sharp forward and backward components of the proton angular distribution are well accounted for by postulating $m_B^{(1)} \approx m_{\pi}$ and $G = 0$. 

Since the proton angular distribution in the ordinary background process is forward peaked, 
we made a strategy to divide observed events according to ``Large Angle Proton" and ``Small Angle Proton" cuts, denoted by LAP with $|{\rm cos} \theta_{\rm cm} (p)|< 0.6$,  and by SAP with $|{\rm cos} \theta_{\rm cm} (p)| > 0.6$, respectively. 
In fact, we have found that the observed Dalitz plots (not shown here) depend very much on the selection of the proton angular distributions \cite{Yamazaki:10}.

\begin{figure}[htb]
\begin{center}
\includegraphics[width=9cm]{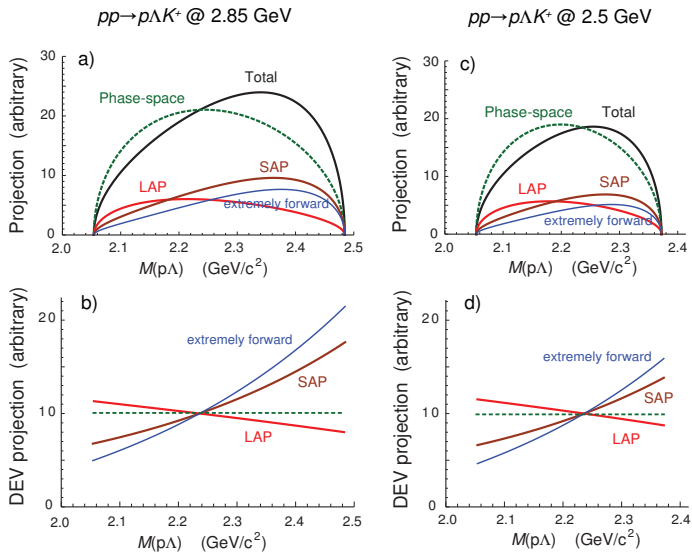}
\caption{(Colour-online) (Upper; a, c) Distributions of $M(p\Lambda)$ (without acceptance correction) for various proton angle groups (LAP, SAP and Total) at $T_p = 2.85$ GeV (a, b) and 2.50 GeV (c, d), calculated for the ordinary three-body process with an intermediate boson mass of $m_B^{(1)} = m_{\pi}$ and $G = 0$ \cite{Akaishi:11}. (Lower; b, d) Corresponding calculated $DEV$ presentations of $M(p\Lambda)$ spectra.}
\label{fig:DEV-ordinary}
\end{center}
\end{figure}

The invariant-mass distributions are then expressed in terms of $T(Q^2)$ by integrating over $X_1$:
\begin{equation}
\frac{d^2 \sigma}{{\rm d}x_{p\Lambda} {\rm d}y_{K\Lambda}} = (\frac{2 \pi}{\hbar c})^6 \frac{E_0}{4 k_0^3} \, \int_{-1}^{+1} |T(Q^2)|^2 {\rm d}X_1, 
\label{eq:Cross-section}\end{equation}
where 
 $x_{p\Lambda} \equiv m_{p \Lambda}^2$ and $y_{K\Lambda} \equiv m_{K \Lambda}^2$.
The $m_{p \Lambda}$ distribution of the Dalitz plot can be calculated by the integration of eq. (\ref{eq:Cross-section}) over $y_{K\Lambda}$. The calculated distributions (without acceptance corrections) and their $DEV$ presentations at $T_p = 2.85$ and 2.50 GeV are shown in Fig.~\ref{fig:DEV-ordinary} for LAP and SAP, as well as for the uniform phase-space. All the projection distributions of $M(p \Lambda)$ (upper figures: a, c) are bell shaped, and thus, not easily distinguishable. On the other hand, their $DEV$ presentations (lower figures; b, d) are nearly linear with easily distinguishable different gradients, which are shown to correspond to different proton angular distributions, reflecting different momentum transfers. Furthermore, for actual experimental data ($RAW$) the $DEV$ distributions are acceptance free, as $SIM$ data take into account the acceptance realistically. The distributions for 2.85 GeV and 2.50 GeV incident energies (shown in  left halves, a, b and right halves, c, d of Fig.~\ref{fig:DEV-ordinary}, respectively) are similar to each other.

The flat large-angle component (LAP) can be explained as the ordinary process (\ref{eq:pLK}) with large $m_B$ values, but it may also involve an exotic two-body process via a $\Lambda^*$ doorway state, 
\begin{eqnarray}
p + p &\rightarrow& \underline{p + \Lambda^*} + K^+, \nonumber\\
                                                 && ~~~~~ \hookrightarrow X  \rightarrow p + \Lambda. 
\label{eq:KX}
\end{eqnarray}                                                 
The existence of such an $X$ can be signaled as a peak in both invariant-mass $M(p\Lambda)$ and missing-mass $\Delta M(K^+)$ $DEV$ spectra.

\begin{figure}[htb]
\begin{center}
\includegraphics[width=9cm]{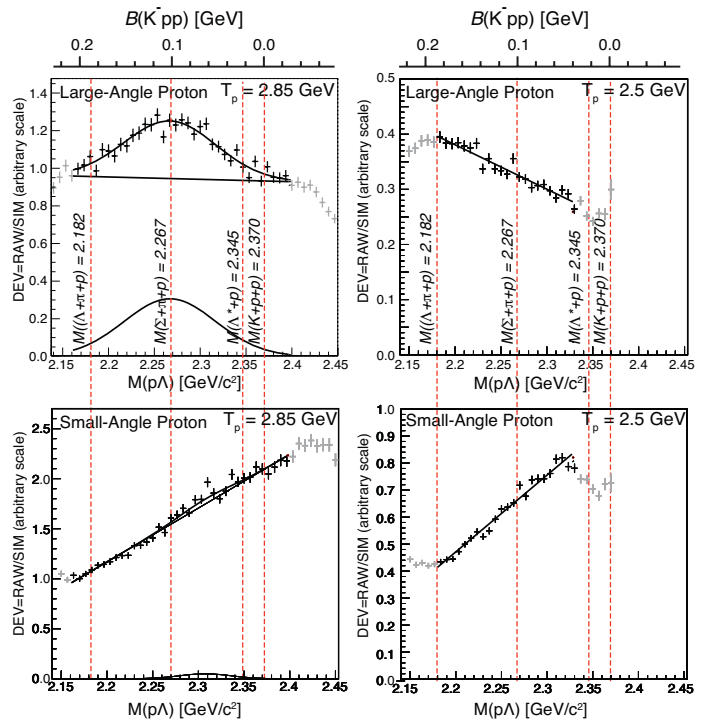}
\caption{Invariant-mass spectra ($DEV = RAW/SIM$ of $M(p\Lambda)$) in arbitrary units for $T_p = 2.85$ GeV (left) and 2.50 GeV (right) incident energies. The upper and lower spectra were obtained by applying Large Angle Proton (LAP) and Small Angle Proton (SAP) cuts, respectively. The thresholds of some relevant decay channels are shown by vertical broken lines. The faint points outside the solid fit zones were discarded because, there, the $DEV$ ratios are not reliable due to the rapidly decreasing acceptance at its boundaries, causing larger systematic errors that cannot be easily assessed.}
\label{fig:MPL_DEV}
\end{center}
\end{figure}

Figure~\ref{fig:MPL_DEV} shows a comparison of the invariant-mass $DEV$ spectra of $M(p \Lambda)$ for $pp$ collisions at incident energies of $T_p$ = 2.85 GeV (left panels) and 2.50 GeV (right panels). The upper spectra at both energies are with LAP cuts, involving a much smaller contribution of the ordinary background, eq.(\ref{eq:pLK}). This selection should not cause any fake effect on the mass spectra, because the proton momentum in c.m. is not so different between the two incident energies; the purpose of the selection is to remove a large amount of extreme forward and backward protons, which are the main source of the background. In fact, the SAP spectra of both incident energies (the lower spectra of Fig.~\ref{fig:MPL_DEV}) show linear behaviors of similar positive gradients without a resonant peak. 
This tendency is the characteristic feature of the ordinary reaction, eq.(\ref{eq:pLK}), when mediated by a low-mass intermediate boson \cite{Akaishi:11}, as shown in Fig.~\ref{fig:DEV-ordinary}, where the calculated distributions and their $DEV$ presentations at $T_p = 2.85$ and 2.50 GeV for $m_B^{(1)} \approx m_{\pi}$ and $G = 0$ are shown for different proton-angle groups as well as for uniform phase-space. 
    
We find a striking difference in the $DEV$ invariant-mass spectra, $M(p \Lambda)$, of LAP between $T_p =$ 2.85 and 2.50 GeV. 
The $M(p \Lambda)$ spectrum at 2.85 GeV shows an outstanding peak that we identified in \cite{Yamazaki:10} as the production of a resonance, $X(2265)$, with high transverse momentum protons in the two-body $p + p \rightarrow K^+ +  X$ reaction followed by $X \rightarrow p + \Lambda$. 
In contrast to this behavior, at 2.50 GeV nearly no trace of the $X(2265)$ contribution is visible. The $M(p \Lambda)$ spectra of both SAP and LAP are totally flat in the mass region of the $X(2265)$ peak; the latter (LAP) shows a negative slope, which is consistent with the simulation given in Fig.~\ref{fig:DEV-ordinary} (c, d), and may also reflect 
a final-state interaction effect between $p$ and $\Lambda$ \cite{COSY-TOF}. To extract the yield, a fit was made with a Gaussian peak, representing the $X(2265)$ process (\ref{eq:KX}) plus a linear background for the three-body process (\ref{eq:pLK}), on the $M(p \Lambda)$ spectra at both incident energies. The $\Delta M (K^+)$ missing-mass spectra show the same behavior as the $M(p \Lambda)$ invariant-mass spectra presented here.

\begin{figure}[tb]
\begin{center}
\includegraphics[width=8.5cm]{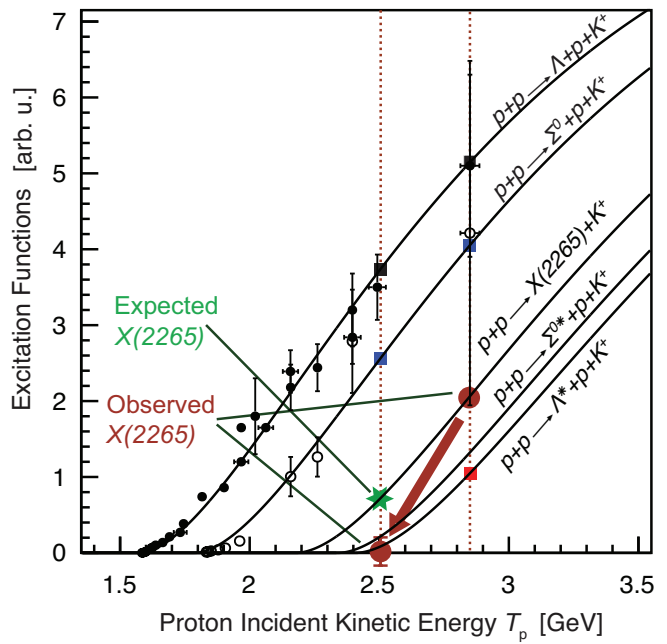}
\caption{{\it RELATIVE} excitation functions in arbitrary units of the reactions $p+p \rightarrow p + \Lambda + K^+$, $\rightarrow p + \Sigma^0 + K^+$, $\rightarrow X(2265) + K^+$, $\rightarrow p + {\Sigma^0}^* + K^+$ and $\rightarrow p + \Lambda^* + K^+$. The curves are drawn by using a universal formula \cite{threshold}, eq.(\ref{eq:universal}), on which known experimental points with error bars of $\Lambda$ (closed circles) and $\Sigma^0$ (open circles) \cite{COSY-TOF} are fitted and located. 
The upper limit of the $\Lambda^*$ production ratio of $T_p$ = 2.50 GeV to 2.85 GeV, $0.10$, derived from Fig.~\ref{fig:DMpK}, is also consistent with the respective curve.
 The observed relative cross sections for $X(2265)$ at 2.50 and 2.85 GeV are shown by large sepia circles, and the expected one at 2.50 GeV relative to that at 2.85 GeV is shown by a green star. The bold sepia arrow indicates the present observation, which is significantly different from the universal curve. }
\label{fig:EnergyDep}
\end{center}
\end{figure}

The yield of the peak $X$ versus the $p \Lambda K^+$ background, defined as
\begin{equation}
Y_X (T_p) = \frac{{\rm Peak~intensity~in}~DEV}{{\rm BG~intensity~in}~DEV}, 
\end{equation}
is estimated to be
\begin{equation}
Y_X(2.85) =  0.168 \pm 0.010,~~
Y_X(2.50) =   0.002 \pm 0.021,   
\end{equation}
and thus the $T_p$ dependence of $Y$ is expressed by the ratio:
\begin{equation}
\frac{Y_X (2.50)}{Y_X (2.85)} = 0.012 \pm 0.125.
\end{equation}
The peak-to-background ratios, $Y_X (T_p)$, are scaled by the cross section $\sigma_{p \Lambda K} (T_p)$ for reaction (\ref{eq:pLK}), which can be derived semi-empirically from the $T_p$ dependence of the $\Lambda$ cross section, as can be seen in Fig.~\ref{fig:EnergyDep}. Then, the ratio of the cross section for $X(2265)$ at 2.50 and 2.85 GeV is obtained as 
\begin{eqnarray} \label{eq:RRX}
R_X^{\rm ~obs} &=& \frac{\sigma_X (2.50)}{\sigma_X (2.85)} = \frac{Y_X (2.50)}{Y_X (2.85)} \times \frac{\sigma_{p\Lambda K} (2.50)}{\sigma_{p\Lambda K}(2.85)} \nonumber \\
&=&  0.009 \pm 0.091, 
\end{eqnarray}
where the value for the $\Lambda$ production cross section ratio of 0.73, obtained from Fig.~\ref{fig:EnergyDep}, is used. To be consistent with the error bar, we consider an upper limit including one standard deviation, that is, 
$R_X^{\rm ~obs} < 0.10$. Note that, despite a possible difference of the detector acceptance at 2.85 and 2.50 GeV, the peak yield, $Y_X(T_p)$, deduced from a $DEV$ spectrum is independent of the acceptance.

To further discuss the implication of this experimental result, we consider the excitation functions ($T_p$ dependence of the {\it RELATIVE} production cross sections) of various strange particles of mass $M$. 
  Figure~\ref{fig:EnergyDep} shows the excitation functions in arbitrary units for the reactions $p + p \rightarrow \Lambda + p + K^+$, $ \rightarrow \Sigma^0 + p + K^+$, $ \rightarrow X(2265) + K^+$, $\rightarrow p + {\Sigma^0}^* + K^+$ and $ \rightarrow \Lambda^* + p +K^+$. They are drawn following a semi-empirical universal form of Sibirtsev \cite{threshold} as a function of the center-of-mass energy ($\sqrt{s}$) common to each with different thresholds  ($\sqrt{s_0} = M + m_p + m_K$), as expressed by
\begin{equation} \label{eq:universal}
\sigma (s) = C \, \sigma_0 \times \left (1-\frac{s_0}{s} \right ) ^{\alpha} \times \left( \frac{s_0}{s} \right ) ^{\beta}
\end{equation}
with two parameters, $\alpha$ and $\beta$, and a constant, $C\, \sigma_0$. It is consistent with what is expected from a simple phase-space dependence. The curves shown are for the best-fit parameters, $\alpha = 1.8$ and $\beta = 1.5$, which we have found using empirical data for $\Lambda$ (closed circles) and $\Sigma^0$ (open circles) productions \cite{COSY-TOF}. 
From these curves one would expect the following ratio for $X(2265)$: 
\begin{equation}
R_X^{\rm ~expected} = \frac{\sigma_X (2.50)}{\sigma_X (2.85)} \approx 0.33,
\end{equation}
{\it if $X$ is an ordinary object that would follow the above relation} (\ref{eq:universal}). 
This is in strong disagreement with the experimental upper limit, $R_X^{\rm ~obs} < 0.10$. 

In summary, we studied the $T_p$ dependence of $X (2265)$ production, and found that the formation cross section at 2.50 GeV is much less than at 2.85 GeV. The origin of this observation may be related to the fact that the formation of a real $\Lambda^*$ resonance drops down at 2.50 GeV. This view is consistent with the proposed picture on the role played by $\Lambda^*$
 as an essential constituent of a kaonic bound state, $K^-pp$ \cite{Yamazaki:02}, and as a doorway particle for the production of $K^-pp$ in $pp$ reactions \cite{Yamazaki:07b}. On the other hand, one might wonder if the presence of nucleon resonances which decay partially to $K^+ \Lambda$, such as $N^*(1650)$ and $N^*(1710)$ \cite{COSY-TOF}, may cause a fake resonance pattern in $M(p \Lambda)$. We actually observe such $N^*$ resonances in DISTO data, but we have confirmed from simulations that their reflections do not produce any fake peak in the $M(p \Lambda)$ distributions. This view is supported by the fact that no peak in $M(p \Lambda)$ is seen at $T_p =$ 2.5 GeV, although the $N^*$ resonances are still observed at the lower  bombarding energy. 
 These aspects will be reported elsewhere in the near future.   
\\
 
We are indebted to a stimulating theoretical discussion with Professor Y. Akaishi. This research was partly supported by the DFG cluster of excellence ``Origin and Structure of the Universe" of Technische Universit\"at M\"unchen and by Grant-in-Aid for Scientific Research of Monbu-Kagakusho of Japan. One of us (T.Y.) acknowledges support by an Award of the Alexander von Humboldt Foundation, Germany.


\end{document}